# REAL-TIME DETECTION OF DEOXYRIBONUCLEIC ACID BASES VIA THEIR NEGATIVE DIFFERENTIAL CONDUCTANCE SIGNATURE

D. Dragoman

Univ. Bucharest, Physics Dept., P.O. Box MG-11, 077125 Bucharest, Romania

M. Dragoman [a],

National Institute for Research and Development in Microtechnology (IMT), P.O. Box 38-160, 023573 Bucharest, Romania

**Abstract**

In this paper we present a method for the real-time detection of the bases of the deoxyribonucleic acid using their signatures in negative differential conductance measurements. The present methods of electronic detection of deoxyribonucleic acid bases are based on a statistical analysis because the electrical currents of the four bases are weak and do not differ significantly from one base to another. In contrast, we analyze a device that combines the accumulated knowledge in nanopore and scanning tunneling detection, and which is able to provide very distinctive electronic signatures for the four bases.

________________________________________________________________________

a) Corresponding author email: mircea.dragoman@imt.ro, mdragoman@yahoo.com

The deoxyribonucleic acid (DNA) molecule is built up of four bases: adenine (A), guanine (G), cytosine (C), and thymine (T), which are attached to sugar and a phosphate backbone. A DNA base together with the sugar and phosphate backbone molecules form a nucleotide. A double-strand DNA is formed by pairing complementary bases (A-G and C-T) of nucleotide sequences. The DNA is the molecule with the highest density of information, which encrypts the genetic code of any living organism, including the humans. Therefore, the decoding of DNA sequence of bases is of paramount importance for understanding life, and for detecting and healing the potential diseases. Presently, the sequencing of a single human genome takes a few months and is a high-cost procedure, exceeding a few millions of dollars. The huge amount of information contained in the human genome is encoded in more than $3 \times 10^9$ base pairs that must be detected and interpreted.

Since the diameter of the DNA helical structure is on the nanometric scale, nanotechnologies are now involved for the electrical detection of DNA bases using several techniques based on DNA translocation through nanopores, which include the measuring of the ionic blockade current through the nanopore, the detection of the transverse nanopore current and the measurement of voltage fluctuations in a capacitor across the nanopore (see the review in Ref. 1). All these detection methods rely on averaging many measurements for the identification of the base sequence since the electronic signals produced by the four bases are weak and quite similar. This low contrast between the electronic signatures of different bases makes their identification difficult and, so, a statistical approach is needed. Solutions to alleviate this problem include the identification of DNA base pairing via the distance decay of the tunneling current between the tip of a scanning tunneling microscope (STM) and the substrate [2], a method that requires more than one functionalized reading head to identify all DNA bases, or the use of an alternating electronic field in a nanopore capacitor [3], which detects the hysteresis of the bases of the DNA strand that moves back-and-forth through the nanopore; such an oscillatory movement increases the detection time.

In what follows, we propose a method to detect in real-time the base sequence of a single-strand DNA molecule that translocates through a nanopore. This method combines the nanopore architecture with the phenomena of field emission and tunneling, specific for STM techniques. The net result is an enhancement of about four orders of magnitudes of the detected current signal from the pA level (specific to the nanopore detection method via the ionic blockade current) up to a few hundreds of nA, and the allocation to each base of a very distinctive pattern of the differential conductance, which jumps from negative to positive values.

The method presented in this paper relies on pulling the single-strand DNA molecule through a nanopore with a diameter of a few nanometers fabricated in a membrane terminated by two sharp electrodes. A dc field is applied transverse to the membrane, which translocates (via the applied force $F$) the DNA through the nanopore due to the fact that the backbone of the DNA is always negatively charged. The two sharp electrodes are used to collect the electrical signal perpendicular to the DNA backbone axis. In contrast to methods based on measurement of the transverse current through the nanopore, the sharp electrodes used in this detection scheme are similar to STM tips and thus are supposed to emit/collect field-emitted electrons that overcome the workfunction of the electrodes and the DNA bases. A schematic representation of the measuring device is illustrated in Fig. 1a.

In this device the electrons are emitted from the sharp electrodes after overcoming its workfunction $\phi$, pass through a generic DNA base B (B can be either A, G, C, or T) that is located in the nanopore at that moment, and are then collected by the other sharp electrode. The electron energy potential distribution that is used to model the device is represented in Fig. 1b, and consists of two potential barriers that represent the vacuum layers that exist between the base B, with diameter $d_B$, and the nanopore edges. The $d_B$ values for all four DNA nucleotides are given in Ref. 1. The nanopore diameter is denoted by $d$ and $\phi_B$ stands for the workfunction

4of base B. So, the DNA base detection proposed in this paper is based on a RTD-type device that works with field-emitted electrons. Since resonant tunneling is involved in the functioning of this device, the resulting current is larger than in usual STM measurements and the electric signatures of the four bases become more dissimilar.

The electrical signal, and in particular the workfunction of the four DNA bases are different and, hence, unique, as follows from the theoretical work in Ref. 4 and from the STM measurements in Ref. 5. Although the electronic signatures of the A, G, C and T bases are quite difficult to calculate, these bases behave, at least when field-emission occurs, as potential barriers with different heights. This fact allows modeling the electron transport through the device in Fig. 1a with the simple energy band configuration depicted in Fig. 1b.

More precisely, in order to extract the $\phi_B$ values from the experimental STM $I$–$V$ characteristics in Fig. 3a in Ref. 5, we have fitted these data with

$$I(V) = V/R + aV^2 \exp(-b/V). \qquad (1)$$

The first term in the right hand side of Eq. (1) represents the contribution of a series resistance $R$ and the last term is a typical Fowler-Nordheim characteristic, which describes field emission from a barrier of height $\phi_B$, which assumes a triangular shape in the presence of an applied electric field. So, DNA bases can be modeled as potential barriers with heights $\phi_B$ for field-emitted electrons, their workfunction being obtained from the expression

$$b = \frac{4L}{3} \frac{\sqrt{2m_0}}{e\hbar} \phi_B^{3/2} \qquad (2)$$

where $L$ is the STM tip-sample distance and $m_0$ the free electron mass. The data in Ref. 5 for all DNA bases could be fitted with series resistances between 500 and 1000 Ω, the average values between the workfunctions obtained from data at positive and negative polarizations being $\phi_A = 1.74$ eV, $\phi_T = 1.81$ eV, $\phi_C = 2.12$ eV, and $\phi_G = 1.7$ eV for an estimated $L = 0.66$ nm.

With these experimentally determined parameters for the four DNA bases, we have modeled the I–V characteristic through the nanopore configuration in Fig. 1 for $d = 2.7$ nm. The results are represented in Fig. 2 with solid line for A, dotted line for C, dashed line for G, and dashed-dotted line for T. The series of peaks that appear in all I–V characteristics indicates the formation of resonant energy levels in the quantum well of the structure represented in Fig. 1b, the quantum well being the DNA base that translocates through the nanopore at the moment the electrical measurement is performed. The DNA bases have different current characteristics due to both different workfunctions and different diameters (different quantum well depths and widths). Since the electrical measurement can be done in a much shorter time than the time it takes for a nucleotide to be pulled through the nanopore (this translocation time is about 1 ms [1]), no confusion about which nucleotides contributes to the signal is possible.

From Fig. 2 it follows that C has a measurable current for significantly lower voltages than the other bases, whereas the current peaks for T, A, and G appear at increasing voltages. Note that a detection method that relies on observing peaks in current rather than differences in current values is potentially more precise. In order to further increase the differences between the electric signatures of the four bases one can measure their differential conductances. The simulations of this parameter, $G$, obtained from the curves in Fig. 1 are illustrated in Fig. 2 with the same line type. As expected, the negative differential conductances take both positive and negative values, the sharp negative peaks of $G$ offering a clear means of identification the four DNA bases.

In conclusion, the differential conductance signature, especially for negative values of the differential conductance, is a sure path in the electronic identification of the four bases of DNA. We have to note that very recently but in a different context, in a molecular nanodevice containing DNA immobilized across gaps formed by gold electrodes, negative differential resistance was found at room temperature [6].

**Figure captions**

Fig. 1  (a) Schematic illustration of the nanopore device configuration for detecting the DNA bases and (b) the energy band distribution used to model it.

Fig. 2  *I–V* characteristics of different bases that translocate through the nanopore: A (solid line), C (dotted line), G (dashed line), and T (dashed-dotted line).

Fig. 3  Differential conductances of the A, C, G, and T bases obtained from the respective characteristics in Fig. 2.

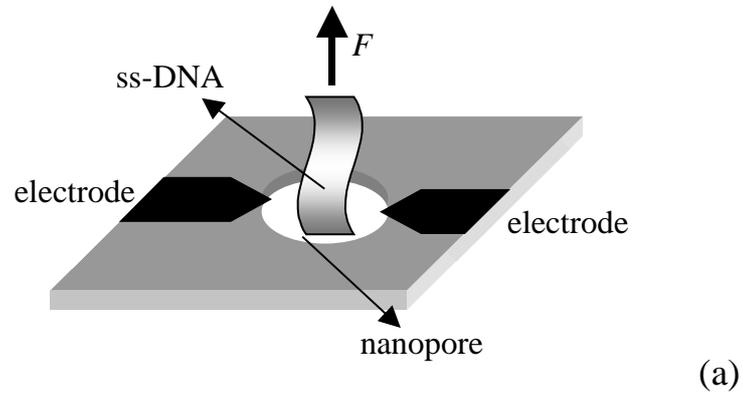

(a)

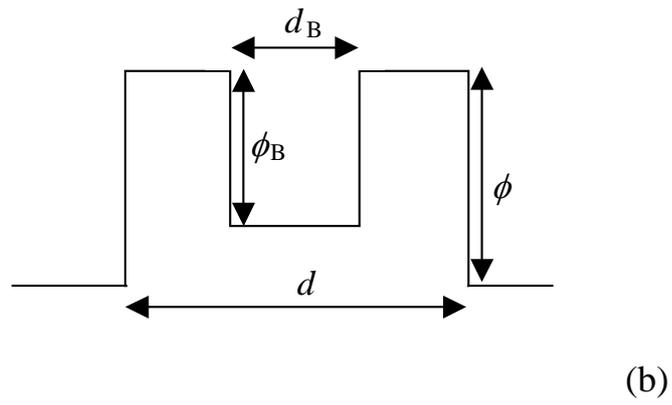

(b)

Fig. 1

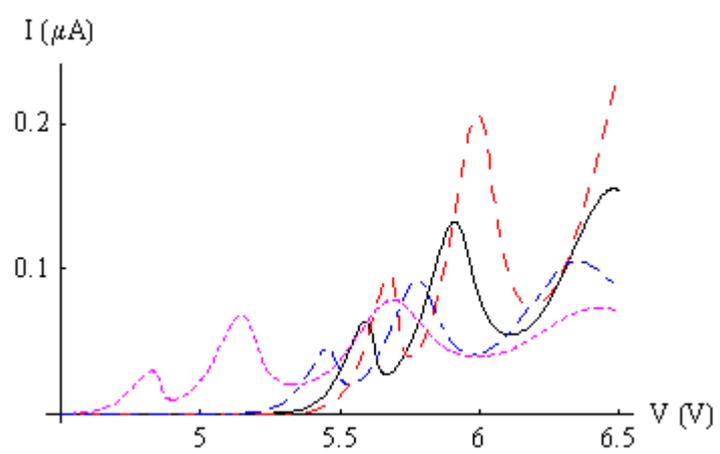

Fig. 2

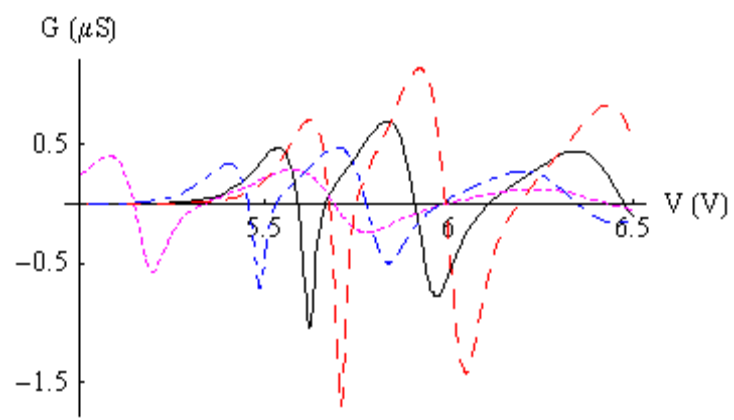

Fig. 3